# A life cycle model for high-speed rail infrastructure: environmental inventories and assessment of the Tours-Bordeaux railway in France.

A. de Bortoli[1*], L. Bouhaya[2], A. Feraille[3]

**1** *Univ. Paris-East, Lab. City Mobility Transportation, Ecole des Ponts ParisTech, Marne-la-Vallée, France*

**2** *Scientific Research Center in Engineering, Lebanese University, Faculty of Engineering I, Tripoli, Lebanon*

**3** *Univ. Paris-East, Lab. Navier Laboratory, Ecole des Ponts ParisTech, Marne la Vallée, France*

*\* Corresponding author : Anne de Bortoli, Université Paris-Est, Laboratoire Ville Mobilité Transports (Ecole des Ponts ParisTech, IFSTTAR, Université Paris-Est Marne-la-Vallée), Ecole des Ponts Paris Tech, 6-8 avenue Blaise Pascal, 77 455 Marne-la-Vallée cedex 2, France. Tel : +33 (0)1 81 66 89 69 / e-mail : anne.de-bortoli@enpc.fr*

## Abstract

*Purpose:* the objective of the study is to progress towards a comprehensive component-based Life Cycle Assessment model with clear and reusable Life Cycle Inventories (LCIs) for High Speed Rail (HSR) infrastructure components, to assess the main environmental impacts of HSR infrastructure over its lifespan, to finally determine environmental hotpots and good practices.

*Method*: a process-based LCA compliant with ISO 14040 and 14044 is performed. Construction stage LCIs rely on data collection conducted with the concessionaire of the HSR line combined with EcoInvent 3.1 inventories. Use and End-of-Life stages LCIs rest on expert feedback scenarios and field data. A set of 13 midpoint indicators is proposed to capture the diversity of the environmental damage: climate change, consumptions of primary energy and non-renewable resources, human toxicity and ecotoxicities, eutrophication, acidification, radioactive and bulk wastes, stratospheric ozone depletion and summer smog.

*Results*: The study shows major contributions to environmental impact from rails (10-71%), roadbed (3-48%), and civil engineering structures (4-28%). More limited impact is noted from ballast (1-22%), building machines (0-17%), sleepers (4-11%), and power supply system (2-12%). The two last components, chairs and fasteners, have negligible impact (max. 1% and 3% of total contributions, respectively). Direct transportation can contribute up to 18% of total impact. The production and maintenance stages contribute roughly equally to environmental deterioration (resp. average of 62% and 59%). Because the End-of-Life (EoL) mainly includes recycling with environmental credit accounted for in our 100:100 approach, this stage has globally a positive impact (-9 to -98%) on all the impact categories except terrestrial ecotoxicity (58%), radioactive waste (11%) and ozone depletion (8%). Contribution analyses show that if concrete production is one of the important contributing process over the construction stage, primary steel production is unquestionably the most important process on all the impact categories over the entire life cycle.

*Conclusion*: These results are of interest for public authorities and the rail industry, in order to consider the full life-cycle impacts of transportation infrastructure in a decision-making process with better understanding and inclusion of the environmental constraints. Suggestions are provided in this way for life cycle good practices – for instance as regards gravel recycling choices – and additional research to reduce the impact of current major contributors.

# 1. Introduction

## 1.1 Context

With the proliferation of threats to human wellbeing and perhaps even survival – e.g. climate change, massive biodiversity loss, air, water, and soil pollution, depletion of resources – it has become increasingly important to question public policies and orientations in the light of environment. The recent profusion of environmental warnings from various international actors – researchers, public and private organizations and even private investment funds – should prompt a more cautious and rational approach.

Transportation is currently one of the major contributors to environmental damage. For instance, it accounts for more than 14% of the anthropogenic GHG emissions worldwide (IPCC 2014, p. 47) and 29% in France (CITEPA 2019), and for a substantial part of the atmospheric pollution: in terms of particulate matter (PM2.5) and nitrogen oxides, transportation would be responsible for respectively 51,000 and 37,000 annual premature deaths in the European Union (European Environment Agency 2018). But its impacts are not always thoroughly taken into account in quantitative terms in public policies. In France, for instance, mandatory cost-benefit analyses are conducted to assess the public benefits of transportation infrastructure investment projects. The quantitative calculation mainly consider two environmental impacts – climate change and air pollution in the operational stage – which are then monetized and aggregated with different financial flows and socioeconomic externalities – namely in France monetized travel times, noise effects on public health, and safety risks (Quinet and al. 2013) – in a single public interest indicator: the SocioEconomic Net Present Value (SE-NPV), calculated according to the equation (1), with $CapEx$ the capital expenditure for the project for the year t=0, $Revenue_{i,t}$ the revenue of type i for the year t, $OpEx_{i,t}$ the operational expenditure of type i for the year t, and $Externalities_{j,t}$ the monetized externality of type j for the year t. Inflation and discount rates must be taken into account before aggregating flows occurring at different years, using well-known formula (Quinet and al. 2013).

$$SE - NPV = -CapEx + \left[\sum_{i,t}(Revenue_{i,t} - OpEx_{i,t}) - \sum_{j,t} Externalities_{j,t}\right]_{discounted+inflated} \quad (1)$$

The environmental gaps in the scope of the life-cycle and impact categories considered in the SE-NPV calculation, and more broadly in the environmental factors used when making decisions on transportation, could be largely alleviated by deploying Life Cycle Assessments (LCA), a multicriteria methodology that has been applied to transportation systems for more than 20 years, and especially on road transportation (Häkkinen and Mäkelä 1996; Horvath and Hendrickson 1998; Ridge 1998; USAMP/LCA 1998). To do so, the French Ministry of transportation would need LCAs of different types of standard infrastructures to allow a fast SE-NPV calculation on a complete life-cycle perimeter. As rail transportation modes are mostly considered amongst the best environmental alternatives within motorized modes considering the only operation stage of transportation (DELOITTE 2008), we propose to assess a railway infrastructure.

Europe has a large railway network, within which France has the second longest with a total of almost 30,000 km (The World Bank 2019). In recent decades, the country has launched a massive High-Speed Rail (HSR) deployment plan, though it has politically slowed down in recent years, due to the scarcity of public funds and the prioritization of transport policies (Zembri and Libourel 2017; Cour des Comptes 2014). In 2019, 2,734 km of HSR are operated in France (International Union of Railways 2019), and a new 302 km-long HSR route has been running between the cities of Tours and Bordeaux since July 2017, reducing passenger time between Paris and Bordeaux from 3 to 2 hours (LISEA 2016). The objective of the study is to assess the environmental impact of this new HSR infrastructure. To do so, we first conducted a review of previous LCAs performed on rail infrastructure.

## 1.2 Background

A dozen railway LCAs have been carried out between 2003 and 2018 (TABLE 1) by researchers and sometimes consultants. System and geographic perimeters, lifespans, data and selected environmental impact categories are some of the factors of variability encountered in these studies. Only few of the studies use field data, which is understandable in the case of ex ante appraisals but less so for ex post ones.

**TABLE 1 Summary of some characteristics of LCA studies performed on rail systems**

| RAIL TYPE | COUNTRY | YEAR | PUBLICATION 1ST AUTHOR | LCI DATA | LINE | SYSTEM PERIMETER | DATA (P/T) | OBJECT (P/T) |
|---|---|---|---|---|---|---|---|---|
| HSR | Germany | 2003 | Rozycki | Railway experts, DB AG | Hanover-Wuerzburg | Mode | P/T | P/T |
| ST | Europe Switzerland | 2004 | Spielmann | Literature | Country | Mode | T | T |
| ST - HSR | Sweden | 2006 | Svensson | Literature | Country | Infra | T | T |
| ST | USA | 2006 | Horvath | Hybrid, literature | USA | Freight modes | T | T |
| ST | Sweden | 2010 | Stripple | Lit., manufacturer data | Nyland-Umea (Bothnia line) | Infra | T/P | P |
| HSR | CA, USA | 2010 | Chester | Hybrid, literature | San Francisco-Anaheim | Mode | T | T |
| HSR | CA, USA | 2011 | Chang | Literature | San Francisco-Anaheim | Infra | T | P |
| HSR | Sweden | 2011 | Akerman | Stripple 2010 | Stockholm-Gothenburg Jönköping-Malmö /Copenhagen | Mode | T | P |
| HSR | Norway | 2011 | Asplan Viak AS | Project, Simapro databases | Oslo-Ski | Infra | P | P |
| HSR | UK | 2012 | Miyoshi | Lit. (Chester, Von Rozycki) | London–Manchester | Mode | T | P |
| ST | Switzerland | 2014 | Fries | EcoInvent v2.2, Shipper survey | Different lines | Freight modes | T | P |
| HSR | China | 2015 | Yue | Chinese Core Life Cycle Database, Ministry | Beijing-Shanghai | Mode | P | P |
| HSR | Turkey | 2015 | Banar and Ozdemir | Literature, EcoInvent | Country | Mode | T | T |
| HSR | Portugal | 2016 | Jones | Literature, EcoInvent v3 | Lisbon-Porto | Mode | T | P |
| HSR | Spain | 2017 | Bueno | Literature | Star-shaped "Y Basque" line | Mode | T | P |

*ST = standard railway; P = Project; T = Theory; Infra = Infrastructure; Lit. = literature US = United States; CA = California;*

Half a dozen academic LCAs have been conducted on specific HSR projects in Germany (Rozycki et al. 2003), California ((Chester and Horvath 2010, 2012; Chang and Kendall 2011)), Sweden (Åkerman 2011), China (Yue et al. 2015), Portugal (Jones et al. 2016) and Spain (Bueno et al. 2017), while non project-specific HSR studies (Svensson 2006; Miyoshi and Givoni 2013; Banar and Özdemir 2015) or non-academic studies (Asplan Viak AS 2011) have also been published. These LCAs mainly use generic data for process-based LCAs or Environmental Input-Output data for hybrid LCAs. To our knowledge, only three studies have been carried out using actual project data: on the German Hanover-Wuerzburg HSR line (Rozycki et al. 2003), the Norwegian Oslo-Ski HSR line (Asplan Viak AS 2011), and the Chinese HSR between Beijing and Shanghai (Yue et al. 2015). Other LCA studies have also been conducted on standard, non-HSR railways ((Spielmann and Scholz 2005; Horvath 2006; Stripple and Uppenberg 2010; Fries and Hellweg 2014)), among which the famous "Bothnia line" LCA using project-specific data (Stripple and Uppenberg 2010). As a conclusion, specific Life Cycle Inventories (LCIs) on HSR infrastructure and feedback on the environmental impacts of non-theoretical projects remain fairly rare.

Among the 3 HSR LCAs based on real project data, Yue et al. (2015) propose the most up-to-date LCIs. But the HSR infrastructure life cycle only includes the construction stage, and excludes all the other stages because of the lack of data on this infrastructure in China. Moreover, the LCIs about the infrastructure construction are specific to HSR in China: design standards can vary widely amongst countries. For instance, the maximum train mass per axle allowed is 12 t in Japan but 15 t in China (Sone 2015), entailing differences in railway design. Or again, due to different safety considerations, double track HSR trace widths are 14 m large in France, Italy or Spain, while they are 15 meter large in the USA, 17 meter large in Belgium or more than 18 m large in Taiwan and Germany (PricewaterhouseCooper 2016). System boundaries in Yue et al. study do not take into account energy systems and signaling system. The quantity of each type of materials is specified in the supplementary data, as well as the lifespans of the infrastructure components: 100 years for engineering structures and roadbed, 20 years for ballast and 15 to 20 years for the railway track (probably referring to rail and sleepers). The lifespans for ballast and railway tracks are especially too short compared to French practices (1.5 to 4 times higher). The study from Asplan Viak AS (2011) proposes a very large system boundary, also taking into account lightening and radio systems. The life cycle is very comprehensive, including the maintenance and EoL stages. The railway lifespan is equal to 60 years, which is very short considering standard railway lifespans as a proxy for HSR lifespan, as HSR are still recent. No information on subsystem lifespans has been provided. Only one kind of concrete has been considered, unlike in Yue et al study (2015). Five impact categories, namely climate change, ozone depletion, acidification, eutrophication and photochemical smog, have been considered. Global material quantities have been given per construction batch, again without consideration of the different physical components (rail, sleepers, etc.). A useful database on building and maintenance machines fuel consumptions is provided. Finally, the first LCA performed on a HSR infrastructure by Rozycki et al. (2000) analyses a German ICE ballastless slab track, different from the traditional gravel bed tracks built in France. It is based on a detailed model considering about 200 items in the inventory. In terms of lifespans, the rail is supposed to last 30 years, the engineering structures 100 years most of the time, and the gravel bed 15 years for gravel and 30 years for concrete and steel elements. Like in the study from Asplan Viak AS, no detailed LCIs are provided, but global quantities of material are given, without any technical specifications, e.g. the type of concrete. Finally, in these three studies, lifespans and/or technologies are different from the French market, and LCIs when provided are not easy to export to create a new HSR model adapted to another context.

The objective of our study is to build a comprehensive model with clear and reusable LCIs for HSR infrastructure components, in order to assess the main environmental impacts of a section of HSR infrastructure over its life cycle in the French case, and to allow practitioners to easily adapt this comprehensive model to conduct HSR infrastructure LCAs in other contexts. To contribute filling the gaps in the environmental assessment of transportation, we performed a LCA on this new section of HSR infrastructure using real construction data supplied by the concessionaire for the line. The study assesses the main environmental impacts resulting from the construction, usage (including maintenance but excluding train operation), and end-of-life of the Tours to Bordeaux HSR infrastructure, and then to discuss environmental levers towards better life cycle management of HSR infrastructure as well as complementary study directions for further research.

## 2. Method

ISO standard 14040 sets out the different steps for performing an LCA followed in this assessment: definition of goal and scope, life cycle inventory collection, impact assessment and interpretation. Throughout this process and at each stage, the analyst needs to check, interpret, and verify the work in progress, in order to fine-tune the assessment and avoid mistakes. The results of the LCA show the potential environmental impacts of the system from cradle to grave, for a given functional unit.

### 2.1 Goal and Scope

The goal of this study is to assess the main contributions to environmental impacts of the Tours-Bordeaux HSR infrastructure section over its life cycle. We choose a process-based attributional LCA method. The functional unit is "allowing high-speed train travel of up to 17 metric tons per axle over a 120-year period" considering French design recommendations, high-speed being a speed of 300 km/h. The section studied comprises 302 km of HSR double track, plus 38 km of standard double track railway connecting to the existing network. Train operation is not included in the scope of assessment. The contribution to environmental impact, then the impacts of the section as well as the impacts for 1 km of single-track over one year, will be provided, in order to allow comparison with other studies. A cross-section of the evaluated railway is shown in FIGURE 1, and describes some of the different subsystems considered (excluding engineering structures): roadbed, sleepers, rails, and the energy system and catenaries. The chairs are plastic pads between the rails and the monobloc steel reinforced concrete sleepers. The fasteners are metal parts made of painted steel that hold the rail to the sleeper.

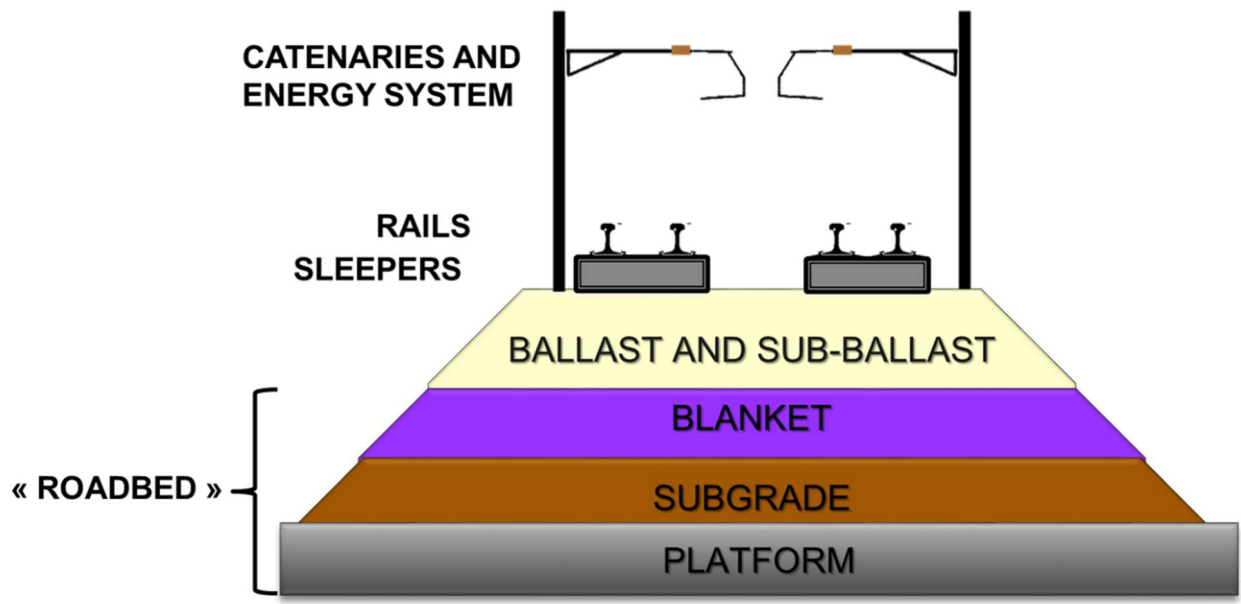

**FIGURE 1  High Speed Rail cross-section**

## 2.2 System Boundaries

The boundaries of the HSR infrastructure system are illustrated in FIGURE 2. We broke the infrastructure down into roadbed, track, power supply system, signaling system, and civil engineering structures. The following are excluded from the scope of the study because of lack of data: maintenance and end-of-life of power supply and signaling systems, roads and drainage, green spaces and fences, excavated material treatment, and bridge bearings. We assume that these elements have limited environmental impacts over the life cycle of the infrastructure.

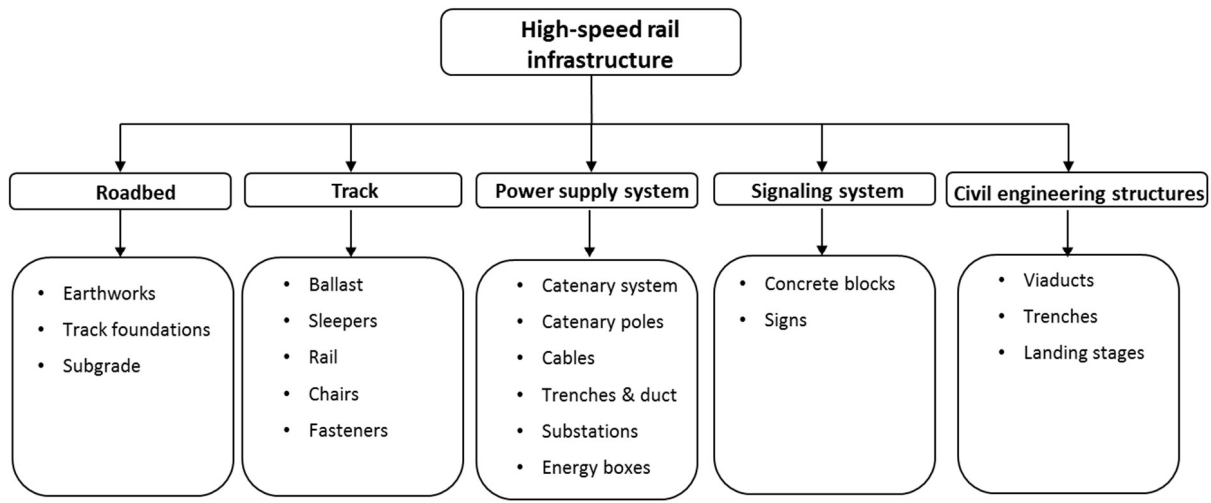

**FIGURE 2** High Speed Rail infrastructure model boundaries with subsystems and components

## 2.3. Lifespan and End-of-Life models

*2.3.1 Lifespans*

Components have different lifespans and possible second lives that need to be considered carefully as well as transparently explained. Assumptions regarding component replacement, maintenance and End-of-Life (EoL) are presented in TABLE 2, based on data provided by French railway experts.

**TABLE 2** Assumptions for replacement, maintenance and End-of-Life of the different components

| COMPONENT | OPERATION | PERIOD (YEAR) | PART | EOL |
|---|---|---|---|---|
| RAIL | Milling | 1 | 100% | N/A |
|  | Replacement | 30 | 100% | 80% recycling - 20% reuse |
| BALLAST | Tamping | 1 | 85% | N/A |
|  | Backfilling | 20 | 15 cm | Recycling |
|  | Replacement | 30 | 30% | Recycling |
| CHAIRS | Replacement | 30 | 100% | Landfill disposal |
| FASTENERS | Replacement | 30 | 100% | Recycling |
| SLEEPERS | Replacement | 60 | 100% | Recycling |

We assume that 20% of the rails are reused on France's secondary rail network after 30 years. The French railway network is classified into 9 subgroups presenting different traffic level intervals, following "International Union of Railways" recommendations. First to fourth class infrastructure, including HSR infrastructure, carries 77% of total French traffic, whereas classes 7 to 9 carry 6% of this traffic. Considering traffic of 100,000 metric tons a day for the first group, 7000 metric tons a day for the second group, and an entire life cycle use stage for one rail of 30 years on HSR plus 30 years on the secondary network, we can allocate the environmental impacts of an object with different uses over its life cycle from the perspective of mechanical wear using generic equation (2), where $M$ is the daily axle load and $D$ the lifespan under this load. According to this calculation, 7% of mechanical rail wear happens on the secondary network.

$$X_i = \frac{M_i D_i}{\sum_j M_j D_j} \qquad (2)$$

All the other steel component parts – sleeper reinforcement steel, fasteners and 80% of the rails – are recycled by producing secondary electric steel, which will be used in other industries instead of primary converter steel. On the basis of the EcoInvent steel production LCI, we assume that 1.105 kg of steel scrap is necessary to produce 1 kg of electric steel. The crushed concrete from the sleepers is recycled as gravel, as well as the ballast, whereas chairs are sent to inert landfill.

*2.3.2 End-of-Life allocation*

There are different methods of EoL allocation, which often attribute very variable environmental impacts to this stage (Nicholson et al. 2009). Among the 11 EoL formula we can find in the literature, we selected a 100:100 approach, and between the 4 choices within this approach, we chose to consider a credit for avoided virgin production at a rate of 100% (Allacker et al. 2017), as this formula is "balanced as it benefits both the products using recycled material and the products designed to be recyclable" (Civancik-Uslu et al. 2019). It means that when recycled materials are used instead of virgin materials at the production stage, we only consider the burden from producing these recycled materials. Based on EcoInvent cut-off system model, we propose to make a correction to add burden and credit from recycling at the EoL according to the formula in equation 3 with $E$ the emissions and resources consumed by the process considered, f including production and EoL stages, v for virgin, $d$ for disposal, R1 the recycled content of material in the input (=0.125/1 for primary steel for instance), R2 the recycled virgin material at the EoL (=1 for steel for instance), Qs and Qp the quality factor of secondary and primary materials (supposed to be =1 for secondary and primary steels) (Allacker et al. 2017):

$$Ef = (1 - R1).Ev + R1.Erecycled + R2.\left(Erecycling, EoL - Ev * \frac{Qs}{Qp}\right) + (1 - R2).Ed \qquad (3)$$

The avoided impact method is encouraged where there is a majority of open-loop recycling, i.e. when the type of recycled materials produced by the system is not massively consumed by that system, and in an attributional LCA with no interaction of the analyzed system with its environment (Le Guern et al. 2011). This is the case here, where the secondary steel for instance is not used in HSR replacement materials. In the avoided impact method, the recycling process for recycled materials is taken into account, as well as the avoided virgin materials production it allows. We then use the 100:100 approach. The whole burdens and avoided burdens from the EoL are attributed to the analyzed system - here the HSR infrastructure - at the EoL stage (and not at the manufacturing stage). Moreover, 100% of the burdens from the production of the virgin material are allocated to the production stage (Civancik-Uslu et al. 2019). It should be noted that no EoL allocation is considered as perfectly representative of the physical reality and that the EoL allocation choice can have an important impact on the final environmental results (Nicholson et al. 2009). We will then present the environmental impacts for each different stage of the life cycle.

## 2.4 Data Collection and Inventory

Data has been collected from several sources. The specific types, quantities and transportation lengths for the construction stage come from construction works reports used internally by the line concessionaire to calculate carbon emissions from construction. These construction works reports are drawn from dozens of different companies involved in the project from preliminary studies to completion. Assumptions for the maintenance and EoL stages come from panels of rail experts from the French National company SNCF and the carbon calculation for an HSR in Eastern France (2009). For generic LCIs, the study relies on EcoInvent V3.1 cut-off system model. Data on the design phase have been ignored. In this section, the different subsystem models are thoroughly presented.

### 2.4.1 Roadbed

Roadbed construction was scheduled in 7 sections, from A to G. Building machine and material consumptions per section were reported to the concessionaire. Data for the entire perimeter is summarized in TABLE 3, which also specifies the EcoInvent V3 processes chosen in the model.

**TABLE 3 Summary of building machine and materials consumption for roadbed construction**

| PROCESS DETAIL | ECOINVENT PROCESS | QUANTITY | UNIT |
|---|---|---|---|
| CONSTRUCTION | diesel, burned in building machines - GLO | 4.01E+09 | MJ |
| MATERIALS | cement, unspecified - RER without CH | 5.07+04 | t |
| | [market, gravel – FR] | 4.21+06 | t |
| | lime, hydraulic - GLO | 2.32+04 | t |
| TRANSPORTATION | transport, freight, lorry 16-32 t, EURO5 - RER | 5.13E+08 | tkm |
| | transport, freight train - FR | 3.62E+07 | tkm |

*GLO = global LCI; RER = Europe; FR = French; CH = Swiss;*

We developed an LCI for the French gravel market. EcoInvent gravel LCIs are based on data from 4 Swiss quarries between 1997 and 2001. France has around 2300 quarries, and the construction sector consumes about 76% of hard-rock and 24% of soft-rock aggregates according to a confidential survey by the French Road Construction Union (USIRF 2016). French LCIs for hard-rock and soft-rock aggregates are not detailed here but are respectively drawn from two survey-based studies by the French Union of Aggregate Producers ((UNPG 2011a, b)), compliant with the French standard NF EN 15804+A1 for the use of LCA in the construction sector (AFNOR 2014) and with the NF EN 14025 relating to environmental labeling. These documents inventory direct input and output flows to produce aggregates from cradle-to-gate, including 42 types of emissions to air, 16 types of emissions to soil, 33 types of emissions to water, 65 flows from natural resources, and 26 flows from the technosphere. Compared to the EcoInvent Unit Process Requirement (UPR) for gravel production that count around two dozen flows in total, these field data collections provide a better completeness, temporal correlation, geographical correlation and further technological correlation for the current French market. Thus, a better quality of LCI after inverse matrix calculation of the final LCI using EcoInvent background dataset, as the number of flows is the UPR is also higher.

*2.4.2 Track*

The track consists of ballast under sleepers, fasteners and chairs, combined with rail, built with a train machine unit.

**TABLE 4 LCIs of each component of the track**

| COMPO-NENT NAME | COMPONENT QUANTITY | PROCESS DETAIL | ECOINVENT PROCESS | PROCESS QUANTITY | UNIT |
|---|---|---|---|---|---|
| **BALLAST** | **1 unit** | Ballast production | [market, gravel – FR] | 2.75E+06 | t/unit |
| | | Building machines | diesel, burned in building machine - GLO | 6.45E+07 | MJ/unit |
| **SLEEPER** | **1.11E+06 units** | Concrete production | concrete, high exacting requirements - GLO | 1.45E-01 | m3/unit |
| | | Prestressed cable manufacture | steel, converter, low-alloyed – RER | 1.25E+01 | kg/unit |
| | | | wire drawing, steel | 1.25E+01 | kg/unit |
| **RAIL** | **1.43E+06 m** | Iron R260 production | steel, converter, alloyed for rail | 6.02E+01 | kg/unit |
| | | Rail manufacture | hot rolling, steel - RER | 6.02E+01 | kg/unit |
| | | Sleeper manufacture | electricity, medium voltage, at grid - FR | 1.51E+01 | MJ/unit |
| | | Rail welding | welding, arc, steel - RER | 5.00E-06 | m/m.unit |
| **FASTENER** | **4.39E+06 units** | Primary steel production | steel, converter, low-alloyed – RER | 3.60 E-01 | kg/unit |
| | | Recycled steel production | steel, electric, low-alloyed – RER | 2.40 E-01 | kg/unit |
| | | Manufacture | hot rolling, steel - RER | 6.00 E-01 | kg/unit |

| CHAIR | 2.22E+06 units | Chair material and manufacture | synthetic rubber - GLO | 4.00 E-01 | kg/unit |
| MACHINES | 1 unit | Train machine unit | diesel, burned in building machine - GLO | 1.69E+08 | MJ/unit |

Machine unit and materials consumptions reported to the concessionaire are shown in TABLE 4 for the 302 km of HSR and the 38 km of standard rail connection. Ballast is then modeled on the basis of the French gravel market LCIs we developed. Engine fuel consumption is modeled using EcoInvent process "*Diesel burned in building machine – GLO*". We developed unitary LCIs for a sleeper, 1 meter of rail, 1 fastener and 1 chair (TABLE 4). TABLE 4 also summarizes how we model the manufacture of each component of the track with specific EcoInvent processes.

*2.4.3 Civil engineering structures*

The main civil engineering structures consist of 6 voussoirs and 4 standard viaducts, 3 landing stages and 1 large covered trench. Construction material quantities reported to the concessionaire are summarized in TABLE 5, as well as the EcoInvent processes chosen in the model.

**TABLE 5  LCI of the civil engineering structures**

|  | PROCESS DETAIL | ECOINVENT PROCESS | QUANTITY | UNIT |
|---|---|---|---|---|
| **CONCRETE MATERIALS** | concrete C30/37 | concrete 30-32 MPa - GLO | 3.46E+04 | m3 |
|  | concrete C25/30 | concrete 25MPa - GLO | 3.64E+04 | m3 |
|  | concrete C20/25 | concrete 20 MPa - GLO | 2.83E+03 | m3 |
|  | concrete C35/45 | concrete 35 MPa - GLO | 1.54E+05 | m3 |
|  | concrete C40/50 | concrete 50MPa - GLO | 2.59E+04 | m3 |
| **REINFORCED STEEL** | reinforced steel | steel, converter, electric - RER | 7.14E+04 | t |
| **WATER SEALING** | bitumen-polymer | bitumen seal, polymer EP4 flame retardant - RER | 4.60E+02 | t |
| **CORNICES** | concrete | concrete, normal - GLO | 1.05E+04 | m3 |
|  | reinforced steel | steel, low alloyed, converter - RER | 8.98E+02 | t |
| **FOUNDATION** | medium capacity stakes | concrete high exacting requirement, GLO | 7.68E+03 | m3 |
|  | high capacity stakes | concrete high exacting requirement, GLO | 4.17E+04 | m3 |
| **STEEL STRUCTURE** | steel | steel, low alloyed, converter - RER | 7.90E+03 | t |

*2.4.4 Construction stage transportation*

Data on transportation during the component manufacturing and the construction stage are given in TABLE 6.

TABLE 6  Construction transportation data

| COMPONENT NAME | ECOINVENT PROCESS | PROCESS QUANTITY | UNIT | SUPPLY DISTANCE |
|---|---|---|---|---|
| BALLAST | transport, freight, rail – FR | 3.57E+08 | tkm | From the concessionaire |
|  | transport, freight, lorry 16-32 metric ton, EURO5 | 5.43E+07 | tkm | From the concessionaire |
| SLEEPERS | transport, freight, rail – FR | 1.18E+06 | tkm | From the concessionaire |
|  | transport, freight, lorry 16-32 metric ton, EURO5 | 9.43E+07 | tkm | From the concessionaire |
| RAIL | transport, freight, rail – RER | 4.21E+01 | tkm | Assumption: 700 km for iron blooms |
|  | transport, freight, rail – FR | 1.51E+01 | tkm | Assumption: 250 km to welding plant and construction sites |
| FASTENER | transport, freight, lorry 16-32 metric ton, EURO5 | 5.27E+05 | tkm | Assumption: 200 km |
| CHAIR | transport, freight, lorry 16-32 metric ton, EURO5 | 6.21E+05 | tkm | Assumption: 700 km |
| CIVIL ENG. STRUCTURES | transport, freight, lorry 16-32 metric ton, EURO5 | 7.60E+07 | tkm | From the concessionaire |

*2.4.5 Power supply and signaling system*

The concessionaire's reporting includes the type of materials consumed for the catenary system, the catenary poles and the connecting cables. A transportation distance of 300 km is assumed for all the elements based on French railway experts' discussions. Data on trenches, energy boxes, and signs are extrapolated based on 1 km of another HSR line in Eastern France for which a carbon calculation had previously been made (2009). The global inventory and EcoInvent choices are indicated in TABLE 7.

TABLE 7 Materials and energy consumption for the construction of the power supply and signaling system -EcoInvent processes chosen in the model

| PROCESS DETAIL | ECOINVENT PROCESS | QUANTITY | UNIT |
|---|---|---|---|
| *TRENCHES* | | | |
| EARTHWORK AND BUILDING MACHINES | Diesel, burned in building machine - GLO | 23115894 | MJ |
| CONCRETE | concrete, normal - GLO | 4046.90022 | t |
| REINFORCED STEEL | steel, converter, low alloyed - RER | 2782.2439 | t |
|  | hot rolling, steel - RER | 2782.2439 | t |
| TRANSPORTATION | transport, freight, lorry 16-32 metric ton, EURO5 | 8.35E+05 | tkm |
| *CATENARY CABLES* | | | |
| CONDUCTING CABLES | cooper - RER | 1.67E+06 | kg |
|  | wire drawing, copper - RER | 1.67E+06 | kg |
|  | steel, low-alloyed, converter - RER | 1.28E+05 | kg |
|  | hot rolling, steel - RER | 1.28E+05 | kg |
| | contouring, bronze - RER | 1.05E+06 | kg |
| **SUPPORTING CABLES** | aluminum, cast alloy - RER | 9.75E+05 | kg |
| **TRANSPORTATION** | transport, freight, lorry 16-32 metric ton, EURO5 | 1.14E+06 | tkm |
| *CATENARY POLES* | | | |
| **STEEL MANUFACTURE** | steel, converter, low alloyed - RER | 14 000 | t |
| | hot rolling, steel - RER | 14000 | t |
| **TRANSPORTATION** | transport, freight, lorry 16-32 metric ton, EURO5 | 7630000 | tkm |
| **CONCRETE BLOCKS** | concrete, high exacting requirements - GLO | 22000 | m3 |
| **TRANSPORTATION** | transport, freight, lorry 16-32 metric ton, EURO5 | 1100000 | tkm |
| **REINFORCED STEEL** | steel, converter, low alloyed - RER | 224 | t |
| | hot rolling, steel - RER | 224 | t |
| **CONCRETE** | concrete, normal - GLO | 893 | m3 |
| | Concrete 25MPa - GLO | 4125 | m3 |
| | concrete 35 MPa - GLO | 2396 | m3 |
| **STEEL STRUCTURE** | steel, converter, low alloyed - RER | 415 | t |
| | hot rolling, steel - RER | 415 | t |
| *CONNECTING CABLES* | | | |
| | cooper - RER | 1.51E+02 | t |
| | wire drawing, copper - RER | 1.51E+02 | t |
| | aluminum, cast alloy - RER | 8.00E+00 | t |
| **TRANSPORTATION** | transport, freight, lorry 16-32 metric ton, EURO5 | 4.77E+04 | tkm |
| *ENERGY BOXES* | | | |
| **STEEL** | steel, converter, low alloyed - RER | 10.1172505 | t |
| | sheet rolling, steel - RER | 10.1172505 | t |
| **TRANSPORTATION** | transport, freight, lorry 16-32 metric ton, EURO5 | 3.04E+03 | tkm |
| *SIGNS* | | | |
| **CONCRETE BLOCKS** | concrete block production - GLO | 442.629711 | m3 |
| **STEEL PANNELS** | steel, converter, low alloyed - RER | 30.3517516 | t |
| | sheet rolling, steel - RER | 30.3517516 | t |
| **TRANSPORTATION** | transport, freight, lorry 16-32 metric ton, EURO5 | 3.41E+05 | tkm |

*2.4.6 Replacement, maintenance and EoL*

The EoL model has been developed based on long-term feedbacks on the first HSR lines in France and on expert interviews from the SNCF, the French national company for railways. Over the infrastructure's 120-year lifespan, considering the component lifespan shown in TABLE 2, rail, chairs, and fasteners are replaced 3 times, and sleepers once. A rapid train-unit machine is used to replace these components, consuming 2000 l/h, used 6 h/day, with a capacity of 600 m/day (single track), for a total fuel consumption of $1.55 \cdot 10^9$ MJ. Rail milling machine consumes 63 MJ/km of simple track. 85% of the ballast is tamped every year, using a 450 kW-powered machine tamping 1 km/h of simple track. Fuel consumption over 2x340 km and 120 years will be $1.12 \cdot 10^8$ MJ of diesel for tamping processes. Track backfilling consists in adding a 15-cm layer of ballast every 20 years: each operation requires 1776.5 t of gravel per kilometer of simple track according to French rail experts, that is to say 4.14 Mt of gravel for the 5 operations on the 340 km of the double track over its 120-year life cycle. 30% of the total ballast mass will also be replaced every 30 years, equivalent to a replacement of 0.9 unit of the quantity of ballast used in construction. Transportation distances are those of the construction stage, i.e. an average of 130 km by train and 20 km by truck. EoL inventories are presented in TABLE 8. The process of production and transportation for 1 m of rail, developed according to the LCIs presented in TABLE 4, is used to model the reuse of rails on the secondary French rail network.

**TABLE 8 Total EoL inventories**

| OPERATION | ECOINVENT PROCESS | QUANTITY | UNIT |
|---|---|---|---|
| *RAIL* | | | |
| RECYCLING | steel, low-alloyed, electric - RER | 2.49E+05 | t |
| | steel, low-alloyed, converter - RER | -2.49E+05 | t |
| REUSING | [rail] | -8.00E+04 | m |
| *SLEEPERS* | | | |
| RECYCLING | steel, low-alloyed, electric - RER | 2.26E+04*(1-0.125) | t |
| | steel, low-alloyed, converter - RER | -2.26E+04 | t |
| DISPOSAL | transport, lorry 16-32t, EURO5 - RER | 2.41E+08 | tkm |
| | waste concrete gravel - CH | 8.04E+05 | t |
| *FASTENERS* | | | |
| RECYCLING | Steel, low-alloyed, electric - RER | 9.54E+03*(1-0.125) | t |
| | steel, low-alloyed, converter - RER | -9.54E+03 | t |
| *CHAIRS* | | | |
| DISPOSAL | transport, lorry 16-32t, EURO5 - RER | 1.06E+06 | tkm |
| | treatment of waste plastic, mixture, sanitary landfill - GLO | 3.55E+03 | t |
| *BALLAST* | | | |
| RECYCLING | gravel crushed - GLO | -4.51E+06 | t |
| | transport, freight train - FR | 5.86E+08 | tkm |
| | transport, freight, lorry 16-32 metric ton, EURO5 - RER | 9.02E+07 | tkm |

## 2.5 Environmental Impact Assessment

This environmental model is implemented in the opensource software OpenLCA 1.4.1.

Since decision-making becomes more complex as the number of indicators rises in an appraisal tool, we selected a dozen indicators encompassing the main environmental impacts of the construction sector in France. The "Cumulative Energy Demand" (CED) to quantify total primary energy demand, the EDIP method to calculate waste productions, and the CML method for all the other impact categories. More specifically, the CED method gives impact factors (and related energy consumptions indicators) for different kinds of energy resources: renewable and non-renewable, from different sources (fossil, nuclear, biomass, etc.). We propose to sum the different energy consumptions to get a global indicator.

# 3 Results

## 3.1 Environmental hotspots

### *3.1.1 Contributors per component*

FIGURE 3 shows the contributions of each component of the section of railway to 13 environmental impact categories over its 120-year life cycle. Rails account for a large majority (10-71% of the impact, respectively minoring and majoring on eutrophication and ecotoxicities), followed by roadbed (2-42%, respectively minoring and majoring on terrestrial ecotoxicity and eutrophication), then the civil engineering structures referred to as "viaducts" (4-28%, respectively minoring and majoring on radioactive waste and human toxicity). We notice more limited contributions from ballast (1-22%, respectively minoring and majoring on freshwater ecotoxicity and eutrophication), building machines (0-16%, respectively minoring and majoring on the same impact categories), sleepers (4-10%, respectively minoring and majoring on human toxicity and climate change) and power supply system (labeled as "catenaries", accounting for 2-12%, respectively minoring and majoring on terrestrial ecotoxicity and bulk waste). The last two components, chairs and fasteners, are almost negligible (respectively max. 1% and 3% of total contributions). Direct transportation contributes from 2% up to 18% of total impact, depending on the indicator.

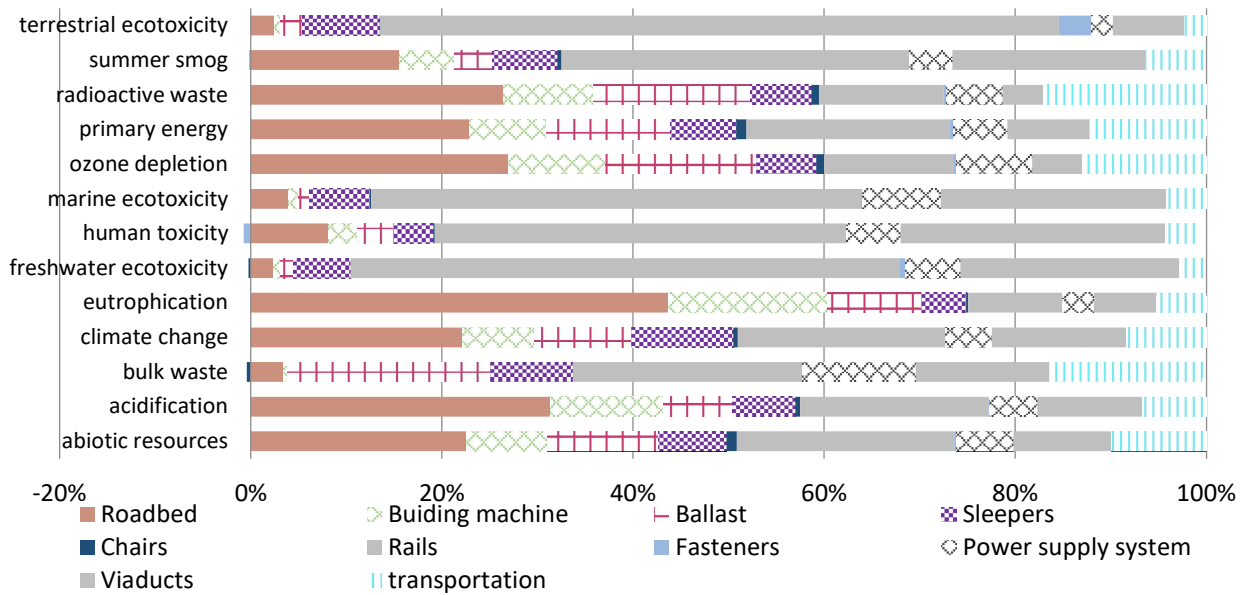

**FIGURE 3** Contributions of each component of the section of railway to environmental impact categories over the 120-year life cycle

*3.1.2 Contribution at each stage of the life cycle*

FIGURE 4 represents the contribution of each stage of the life cycle to the different impact categories. Because our EoL scenario mainly includes recycling, this stage has a positive global impact (-9 to -98%, respectively minoring and majoring on primary energy and human, toxicity) on all the impact categories except terrestrial ecotoxicity (58%), radioactive waste (11%) and ozone depletion (8%). The production and maintenance stages contribute roughly equally to environmental deterioration (resp. average of 62% and 59% while the End-of-Life contributes in average to -21% of the impacts).

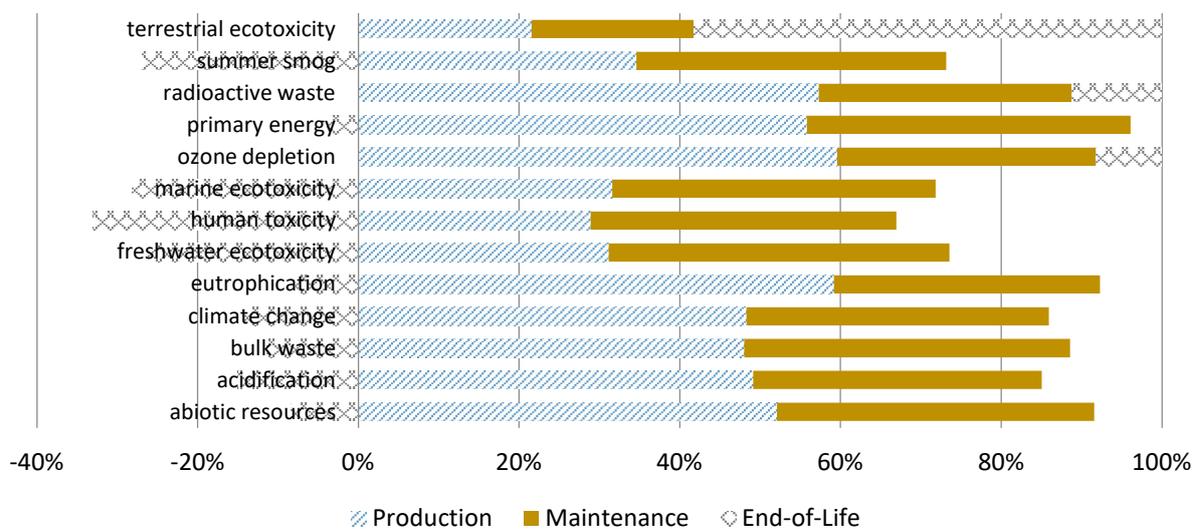

**FIGURE 4  Contribution of each life-cycle stage to the different impact categories**

*3.1.3 Construction stage*

FIGURE 5 details the contribution of each subsystem to the different impacts during the construction stage: the roadbed subsystem is the biggest contributor, with 5 to 66% of total impact – respectively minoring and majoring on freshwater ecotoxicity and eutrophication – and an average contribution of 34%, followed by rails (7%-41%-22%, respectively minoring and majoring on eutrophication and ecotoxicities), civil engineering structures (9%-36%-22%, respectively minoring and majoring on radioactive waste and human toxicity), power supply systems (noted "catenaries and signage", accounting for 5%-19%-10%, respectively minoring and majoring on eutrophication and bulk waste), sleepers (4%-9%-7%, respectively minoring and majoring on eutrophication and climate change), and ballast (1%-11%-5%, respectively minoring and majoring on freshwater ecotoxicity and radioactive waste). Chairs, fasteners, and track building machines are almost negligible, with average impacts of less than 1%.

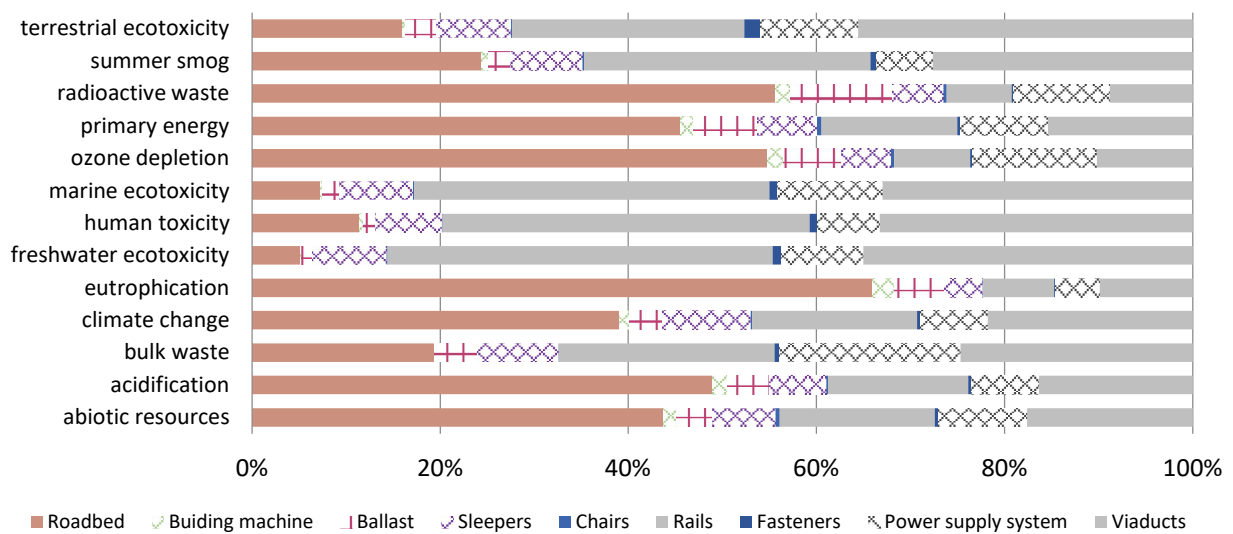

**FIGURE 5** Contribution of each subsystem to the different impacts during the construction stage

In the construction stage, three groups stand out as main contributors. Steel production then transportation by trucks are the two main contributors for abiotic resources depletion (respectively contributing to 57% and 14% of total impact), bulk waste production (resp. 63% and 19%), human toxicity (resp. 92% and 3%), ozone depletion (resp. 37% and 26%), primary energy consumption (resp. 52% and 15%) and radioactive waste production (resp. 29% and 26%). Steel production then concrete manufacturing are the principal sources of acidification (resp. 60% and 15%), climate change (resp. 57% and 23%), eutrophication (resp. 46% and 10%), summer smog (resp. 76% and 6%) and terrestrial ecotoxicity (resp. 52% and 19%). Steel production then copper production are the main contributors regarding freshwater ecotoxicity (resp. 87% and 2%) and marine ecotoxicity (resp. 82% and 3%). The steel production contributor includes both primary and secondary steels, but the impact of secondary steel accounts for few percent of the impact among the environmental categories. The concrete production also aggregates all the different qualities of concrete. Nevertheless, the high-quality concrete ("high exacting requirements") then the 35 MPa concrete are presenting the main contributions.

*3.1.4 Maintenance stage*

FIGURE 6 presents the contribution of each subsystem to the different impacts during the railway's operating stage. Rail is undoubtedly the main contributor to impacts in this stage, with an average contribution of 67% (33 to 90%, respectively minoring and majoring on radioactive waste and freshwater ecotoxicity). The second biggest contribution comes from the use of maintenance machines, accounting for an average of 14% of the impacts (1 to 38%, respectively minoring and majoring on bulk waste/fresh water ecotoxicity and eutrophication). Transportation, ballast maintenance, and sleeper replacement contribute to around 6% of the environmental burdens. Chairs and fasteners are negligible.

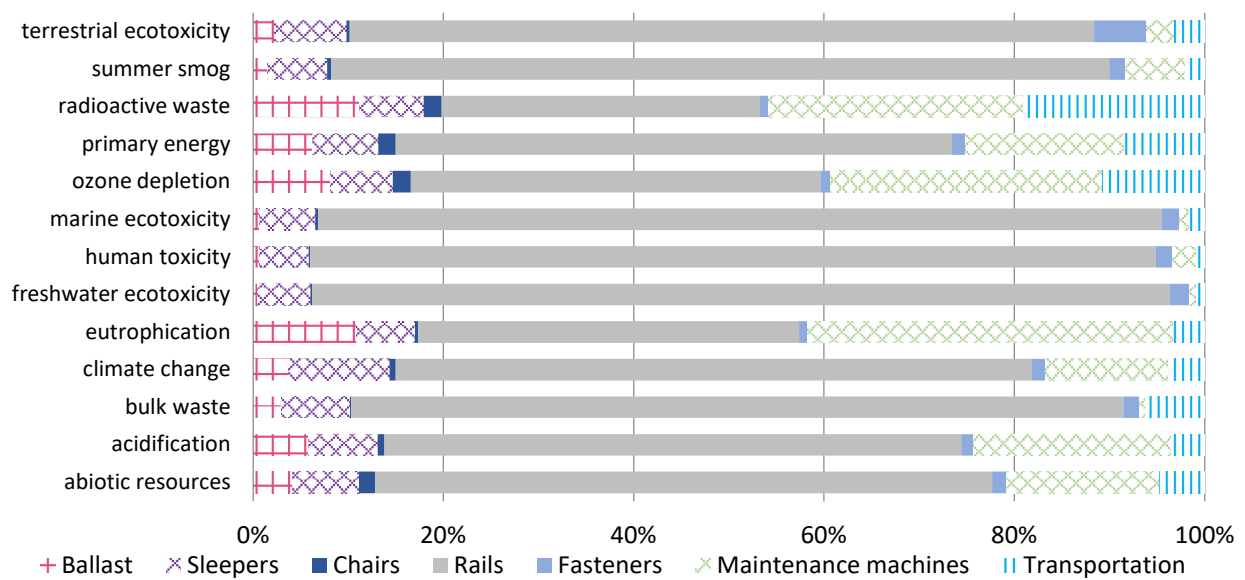

**FIGURE 6  Contribution of each subsystem to the different impacts during the maintenance stage**

In the maintenance stage, the steel production process is the most impacting process in all the impact categories but eutrophication, where it is only the second most contributing process (accounting for 39% of total impact), and radioactive waste production that is mainly impacted by freight by train (29% of total impact) due to electricity consumption of the trains in France (the French electricity mix comes at 70% from nuclear plants), and by diesel burned in building machines ( 24% of total impact). The steel production process accounts for around 90% of three toxicities (freshwater and marine ecosystems, as well as human), 80% of summer smog, 75% of bulk waste production and terrestrial ecotoxicity, 60% of climate change, 55% of abiotic resources depletion and acidification, 45% of primary energy consumption, 35% of eutrophication, and 30% of ozone depletion. The diesel burned in the building machines used to maintain the railway is the first contributor of eutrophication ( 39%) and the second of ozone depletion (28%, steel hot rolling being almost as impacting), acidification (22%), abiotic resources depletion and primary energy consumption (17%), climate change (14%), summer smog (7%) and human toxicity (3%). Finally, the hot rolling of steel is also the second most impacting flow in freshwater and marine ecotoxicity categories (resp. 5% and 4%), while freight by truck is the second most impact flow in the waste production category (6%).

*3.1.5 End-of-life stage*

The EoL environmental impacts of the different track components are illustrated in FIGURE 7. It highlights the environmental benefit of producing secondary steel from primary steel scrap in order to reduce virgin steel production: the principal source of steel scrap is rail, but the positive effects of recycling steel from fasteners and sleepers is not negligible (in blue). The benefit is more mitigated with regard to radioactive waste, as the production of electrical steel consumes 20 times more electricity than the production of primary steel. Ballast recycling has a major negative impact on the environment, since it must be transported by truck over 300 km, while the impacts of producing virgin gravel are relatively small.

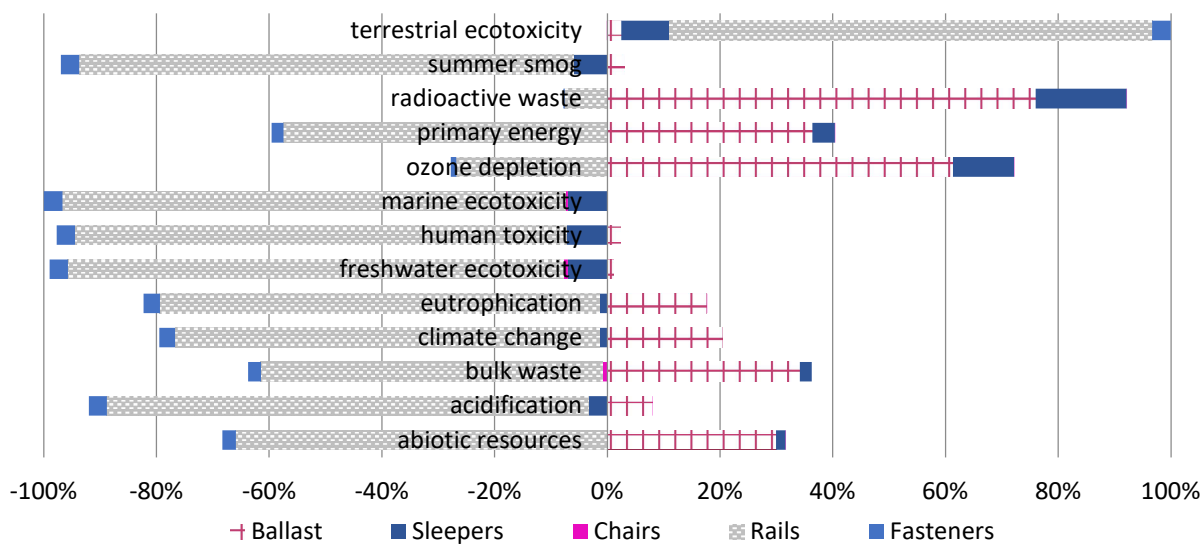

**FIGURE 7** Contribution of each component to the different impacts during the EoL stage

In the EoL stage, the two main contributors are almost homogeneous amongst impact categories: they are first incineration of municipal solid waste (for all categories), happening in the recycling of steel, and then freight by truck. Only freshwater and terrestrial ecotoxicity categories are secondly more impacted by the production of recycled steel. Nevertheless, the production of pig iron from recycled steel, and then the use of former high-speed rail on low-traffic railways, are important processes offsetting part of the burden in many categories. These results are of course related to the EoL methodological choices made in this study. The 100:100 approach is sometimes criticized for double counting recycling benefits, i.e. at the material consumption stage and at the EoL: if this is true for a global environmental accounting, this approach is nevertheless totally valid for a system narrow enough, as far as the transparency of the methodological choices is ensured.

### 3.2 Quantitative impacts

TABLE 9 shows the average environmental impacts of the railway over its life cycle per km and per year, for a single track.

**TABLE 9 Quantitative environmental impacts of a single track life cycle per kilometer and year**

|  | ROAD-BED | MA-CHI-NES | BAL-LAST | SLEE-PERS | RA-ILS | CHA-IRS & FAST. | ENER-GY SYST. | VIA-DUCTS | TRANS-PORT | TOT. |
|---|---|---|---|---|---|---|---|---|---|---|
| **RESOURCES (KG ANTIMONY-EQ)** | 3.63 E+01 | 1.36 E+01 | 2.11 E+01 | 1.14 E+01 | 3.65 E+01 | 2.15 E+00 | 9.51 E+00 | 1.66 E+01 | 1.81 E+01 | 1.67 E+02 |
| ACIDIFICATION (KG SO2-EQ) | 3.97 E+01 | 1.49 E+01 | 1.21 E+01 | 8.37 E+00 | 2.49 E+01 | 7.13 E-01 | 6.38 E+00 | 1.40 E+01 | 9.98 E+00 | 1.32 E+02 |
| BULK WASTE (KG) | 3.01 E+02 | 3.41 E+01 | 2.06 E+03 | 7.59 E+02 | 2.12 E+03 | -3.73 E+01 | 1.05 E+03 | 1.23 E+03 | 1.57 E+03 | 9.06 E+03 |
| CLIM. CHANGE (KG CO2-EQ) | 5.90 E+03 | 2.00 E+03 | 3.14 E+03 | 2.83 E+03 | 5.79 E+03 | 1.39 E+02 | 1.28 E+03 | 3.74 E+03 | 2.57 E+03 | 2.75 E+04 |
| EUTROPHICATION (KG NOX-EQ) | 7.20 E+01 | 2.75 E+01 | 2.17 E+01 | 7.70 E+00 | 1.63 E+01 | 3.55 E-01 | 5.43 E+00 | 1.08 E+01 | 9.96 E+00 | 1.72 E+02 |
| HUMAN TOXICITY (KG 1,4-DCB-EQ) | 3.45 E+03 | 1.31 E+03 | 1.83 E+03 | 1.79 E+03 | 1.83 E+04 | 5.51 E+01 | 2.38 E+03 | 1.18 E+04 | 1.78 E+03 | 4.20 E+04 |
| FRESHWATER ECOTOXICITY (ID) | 3.56 E+02 | 1.02 E+02 | 2.54 E+02 | 8.84 E+02 | 8.44 E+03 | -2.71 E+02 | 8.41 E+02 | 3.36 E+03 | 4.71 E+02 | 1.48 E+04 |
| MARINE ECOTOXICITY (ID) | 1.03 E+06 | 2.89 E+05 | 4.72 E+05 | 1.66 E+06 | 1.35 E+07 | 8.71 E+04 | 2.14 E+06 | 6.21 E+06 | 1.37 E+06 | 2.68 E+07 |
| OZONE DEPLET. (KG CFC-11-EQ) | 9.43 E-04 | 3.57 E-04 | 6.24 E-04 | 2.19 E-04 | 4.76 E-04 | 3.87 E-05 | 2.77 E-04 | 1.83 E-04 | 5.37 E-04 | 3.69 E-03 |
| PRIMARY ENERGY (MJEQ) | 8.61 E+04 | 3.06 E+04 | 5.82 E+04 | 2.60 E+04 | 8.03 E+04 | 5.32 E+03 | 2.12 E+04 | 3.25 E+04 | 5.60 E+04 | 4.02 E+05 |
| RADIOACTIVE WASTE (KG) | 5.67 E-01 | 2.01 E-01 | 4.25 E-01 | 1.37 E-01 | 2.79 E-01 | 2.46 E-02 | 1.26 E-01 | 9.22 E-02 | 4.65 E-01 | 2.34 E+00 |
| SMOG (KG ETHYLENE-EQ) | 1.06 E+00 | 3.94 E-01 | 3.33 E-01 | 4.59 E-01 | 2.48 E+00 | 2.08 E-02 | 3.04 E-01 | 1.37 E+00 | 5.01 E-01 | 6.94 E+00 |
| TERRESTRIAL ECOTOXICITY (KG 1,4-DCB-EQ) | 7.75 E+00 | 2.07 E+00 | 8.66 E+00 | 2.50 E+01 | 2.20 E+02 | 1.09 E+01 | 6.95 E+00 | 2.34 E+01 | 8.39 E+00 | 3.24 E+02 |

The quantitative impacts of the railway are compared to two different processes from the EcoInvent database: "*Railway track construction, for high-speed train*" in a German context (DE) and in the rest of the world (RoW) (Spielmann et al. 2007). The EcoInvent LCIs are valid for the period 2000-2014. Despite the name of the processes, they include a maintenance stage as well as some disposal at the EoL. These processes refer to a ballastless double track over its 100-year lifespan with 15% of trenched tunnels, 23% of mined tunnels, 8% of rail glen bridges and 1% of railroad bridges on the total track length. The result of this comparison is presented on FIGURE 8, with impacts normalized by the EcoInvent process for the rest of the world context. Impacts calculated for the Tours-Bordeaux infrastructure are globally lower than the reference: from -7% on the terrestrial ecotoxicity impact category to -76% on the bulk waste category, with an average of -37%. Even though the models present several differences (lifespan, system boundaries, region, etc.), we infer these differences could mainly be explained by the important length of viaducts, tunnels and trenches structures in the EcoInvent models, compared to the French case study that includes few trenches and viaducts, and no tunnel. Indeed, Chang and Kendall has shown that 15% length of tunnels and aerial structures contribute 60% of greenhouse gas emissions of the California's HSR construction (2011).

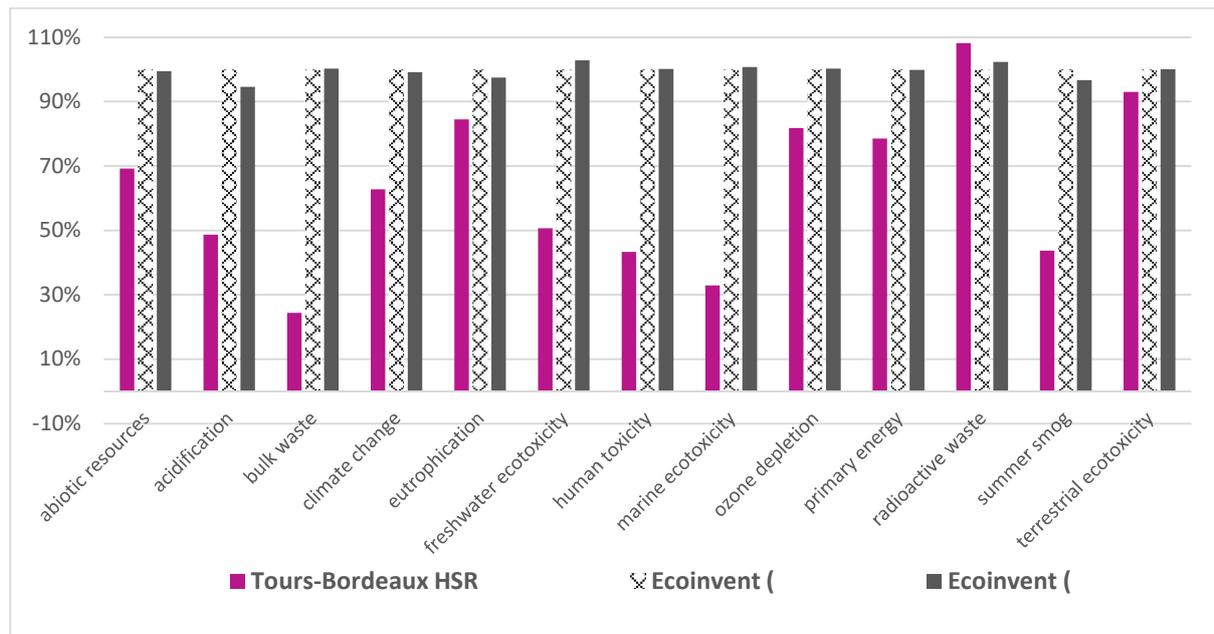

**FIGURE 8** Comparison of the environmental impacts of a double track high speed rail infrastructure over its entire lifecycle, per kilometer and per year

The EoL model must also explain the result differences – and for instance the cut-off allocation in EcoInvent that does not count credits for recycling at the EoL Vs our 100:100 allocation that does, but the dedicated EcoInvent report is not thoroughly detailed (Spielmann et al 2007) - as well as the system boundaries. The (i) use and emissions to soil of lubricants and herbicides for weed control, as well as land (ii) transformation and occupation are accounted for in EcoInvent but not in this study for two reasons: lack of data for (i) or uselessness considering the set of impact categories (ii). We did not want to reuse EcoInvent LCIs as they have been developed two decades ago when practices were different: for instance, between 2001 and 2012, we estimate that the use of phytosanitary products on French railways has decreased by 54% (Senat and SNCF - Service environnement 2001; Manche-nature 2016). On the contrary, our perimeter includes signaling infrastructure while it is not the case in EcoInvent. Different lifespans, shorter than 100 and 120 years respectively, are considered for the different components of the HSR infrastructure in both EcoInvent models and this study: as we compare annual impacts, only the roadbed and potential other components that last the entire lifespan must make vary the final impacts. The EcoInvent LCIs are also representative of the construction, maintenance and EoL in 2000, which must have changed consequently (Spielmann et al. 2007). All these reasons may explain the smaller impacts found in our model. Only the radioactive waste impact is higher for the Tours-Bordeaux infrastructure than for the EcoInvent reference, which is coherent with the differences between the French electricity mix, mainly from nuclear energy, and the "average" world and German electricity mixes.

# 4 Discussion

## 4.1 LCA against environmental burden shifting and "systematic" circular economy practices?

This study shines the spotlight on the importance of end-of-life management and modeling choices. At an overall system level, the 100:100 allocation with credits from the use of recycled materials instead of virgin materials leads to a double count of the burden from recycling. Nevertheless, at the HSR project level, this methodological choice is physically realistic and favors a circular economy in promoting both the use of recycled materials and the recyclability of the system. With the avoided impact method and a 100:100 approach, recycling steel scrap into secondary steel has a massive impact on HSR's life-cycle environmental account. It offsets part of the environmental burden for 12 indicators but shifts it to terrestrial ecotoxicity. It thus casts light on the question of circular economy which always favors a closed-loop option, but rarely from a systems perspective among practitioners (Kirchherr et al. 2017). The example of ballast recycling in this case study shows the limitations of this approach. Indeed, in this study, recycling ballast is not environmentally friendly because of important transportation distances covered by trucks. It would obviously be beneficial to the environment to reuse it on site, but this possibility would be reliant on local demand and supply as well as the need for gravels to be washed in special centers or not. On the basis of the LCA impacts of different ballast freight modes, it would be possible to look for transportation solutions in which ballast recycling would actually be environmentally beneficial, or if no option is satisfactory, to wait for a local demand. More generally, LCA can be used as a support tool to check the objective effectiveness of purportedly pro-environmental practices.

## 4.2 LCA on a transportation mode perimeter

Taking environmental impacts into consideration when making decisions on the construction, operation, and EoL of public infrastructure, contributes to a global move towards sustainability. Nevertheless, the function of a transportation infrastructure is to provide mobility for vehicles, and finally for passengers and goods. The aim of this study was to thoroughly assess the infrastructure of the HSR, excluding the trains running on the line. The model and results can be used as a submodel when assessing a HSR transportation mode or as a base to be adapted in other contexts. Over the global life cycle of a transportation mode, the contribution of the infrastructure subsystem to the total environmental impact varies, depending on several variables like the mode type, technical characteristics, traffic and vehicle occupancy, geographic scale, the impact categories, etc. According to LCAs conducted on HSR modes, infrastructure construction and maintenance are important and could account for more than half of the environmental impact on indicators such as climate change, energy consumption, or air pollution, when considering future technical improvements to trains (Chester and Horvath 2012). Not only is the HSR infrastructure environmentally important in itself, but the energy consumption of trains partly depends on it: there are obvious environmental CapEx-OpEx tradeoffs between infrastructure geometry and train operation costs (Bosquet et al. 2014), depending on speed.

### 4.3 Traceability, regionalization and decision-making

The issue of component traceability to reduce uncertainties in LCIs and eventually on LCA results is particularly important in this respect, especially the traceability of the rail that has a major impact on the HSR infrastructure environmental performance.

The LCA conducted here is based on static LCIs. Using scenarios must improve the quality of transportation LCA to tackle sociotechnical uncertainties, for instance inevitable future transformations in the energy sector as well as in transportation vehicles and practices. Different scenarios need to be considered in order to explore the role of HSR transportation modes in our society ((Åkerman 2011; Perl and Goetz 2015)), as part of a multimodal network, leading to the classic question of competition between short-haul flights and HSR modes from an environmental perspective ((Chester and Ryerson 2014; Albalate et al. 2015; D'Alfonso et al. 2015)), but also to the more complex considerations of network effects on extra traffic demand and road modal shifts (Åkerman 2011).

Finally, while traceability and the regionalization of LCIs may lead to more accurate environmental assessment thus better pro-environmental decisions, they may also cause hidden socio-economic costs, such as the relocation of production outside the country to artificially reduce environmental impact inside the country, with important side effects on employment and its social conditions.

### 4.4. Uncertainties

As uncertainties occur at every phase of an LCA model (Baker and Lepech 2009), performing a sensitivity analysis is a recommended practice to test or generalize LCA results over a larger range of similar kind of objects, e.g. HSR across a territory (region, country, larger area). As performing a sensitivity analysis is time-consuming over the already large effort to get regional LCIs in the previous inventory phase, a common practice across the LCA community is to use a one-at-a-time (OAT) parameter sensitivity analysis, a straight parametric sensitivity analysis changing the value of one parameter before making the model run again. Nevertheless, some specialists consider this as an inappropriate practice that must be avoided, to prefer Monte Carlo simulations or global sensitivity analysis methods (Saltelli and Annoni 2010). Such a work could be the object of a second study on HSR infrastructure. Our study especially highlights the environmental importance of the construction of roadbed and civil engineering structures, the rail production, the maintenance scheme as well as the EoL choice. Rate over one kilometer of civil engineering structures (both for tunnels and bridges), railway material quantities per kilometer, component lifespans and recycling rates could be some of the parameters to be tested in an uncertainty study, along with other classical parameters such as the electricity mix that is an obvious source of variability and uncertainty.

## 5. Conclusion

HSR construction and maintenance data are scarce resources that potentially vary with each country's standards, depending on many variables such as train maximum speed or mass per axle, or again track width and design safety construction coefficient. These data are thoroughly detailed in this first Life Cycle Assessment for a French HSR. Indeed, among the HSR LCAs conducted around the world, only a few are project-specific, since most such studies are prospective. Here, detailed inventories of the track components and their respective lifespan and maintenance processes are provided. These data will be available for future HSR and standard railway assessments, as well as for the eco-design of individual components and materials.

This very detailed process-based LCA of a section of HSR infrastructure, using ex-post building data, shows the respective environmental burdens of its different subsystems over its entire life cycle. Contribution analyses also show that if concrete production is one of the important contributing process over the construction stage, primary steel production is unquestionably the most important process on all the impact categories on the entire life cycle. The identification of environmental hotspots – rail, construction of roadbed and civil engineering structures, EoL choices – could provide new perspectives for the direction of public and private research and development (e.g. targeting life cycle enhancements in the concrete and steel industries, including a life cycle perspective), project decisions (geographical location of infrastructure depending on topography and geotechnical qualities), as well as public policies (technical infrastructure choices, channels for recycling and reuse). These results are a first estimate of HSR infrastructure impacts over its life cycle: they can also be used in the mandatory CBA in France to complete its current assessment scope for HSR projects.

**AKNOWLEDGEMENTS**

This work was funded by the chair ParisTech-Vinci "Eco-design of buildings and infrastructures", a five-year collaborative research program started in 2008 and renewed in 2013, conducted by Ecole des Ponts ParisTech, Mines ParisTech and AgroParisTech thanks to the sponsorship of VINCI. We would especially like to express our gratitude to LISEA, the concessionaire of the HSR line, which helped us to collect the data relating to the construction works.